\documentclass[twocolumn,showpacs,aps,prl,letterpaper]{revtex4}
\usepackage{graphicx}
\usepackage{dcolumn}
\usepackage{amsmath}
\usepackage{epsfig}
\long\def\inst#1{\par\nobreak\kern 4pt\nobreak
    {\itshape #1}\par\vskip 10pt plus 3pt minus 3pt}
\RequirePackage{xspace}
\usepackage{relsize}

\begin{document}

\title{\large \bfseries \boldmath Wave functions and decay constants of $B$ and $D$
mesons\\
in the relativistic potential model}
\author{Mao-Zhi Yang}\email{yangmz@nankai.edu.cn}
\affiliation{School of Physics, Nankai University, Tianjin 300071,
P.R. China}

\date{\today}
%\date{March 32, 2003}

\begin{abstract}
With the decay constants of $D$ and $D_s$ mesons measured in experiment
recently,  we revisit the study of the bound states of quark and antiquark in
$B$ and $D$ mesons in the relativistic potential model. The relativistic
bound state wave equation is solved numerically. The masses, decay constants and
wave functions of $B$ and $D$ mesons are obtained. Both the masses and decay
constants obtained here can be consistent with the experimental data. The wave functions can be
used in the study of $B$ and $D$ meson decays.
\end{abstract}

\pacs{12.39.Pn, 14.40.Lb, 14.40.Nd}
% PACS, the Physics and Astronomy Classification Scheme.

\maketitle

\section*{I Introduction}
The wave function of the bound state of quark and antiquark is
determined by the strong interaction between quark and antiquarks.
The study of the wave functions of heavy-flavored mesons like $B$
and $D$ are important not only for studying the property of strong
interaction between heavy and light quarks, but also for
investigating the mechanism of heavy meson decays. The wave
function determines the momentum distributions of the quark and
antiquark in mesons, which is an important quantity for
calculating the amplitude of heavy meson decays \cite{QCDF,PQCD}.
The light-cone momentum distribution amplitudes for $B$ meson
appear in the amplitude of $B$ decays, which are defined through
the hadron-to-vacuum matrix element of non-local operators of
quark and antiquark separated along the light-cone $\langle 0|
\bar{q}^\beta (z)b^\alpha (0)|\bar{B}(p)\rangle $. The light-cone
distribution amplitudes of $B$ meson have been extensively studied
in the recent several years. Some properties of the
light-cone distribution amplitudes have been
obtained. Based on these achievements, several models satisfying
these constraints have been proposed in the literature
\cite{B1,EOM,EOM2,QSR,KtF,ntwist,Hw}. Methods to obtain the
light-cone distribution amplitude exactly from the first principle
of QCD are still under investigation.

Alternatively, directly studying the wave functions of the heavy
mesons by solving the bound state wave equation is an effective
way to obtain the knowledge about the bound state of quark and
antiquark \cite{RGG,GI,Ce,CCCN,PNP,CF}. For a heavy-light system,
in the heavy quark limit, the heavy quark can be viewed as a
static color source in the rest frame of the heavy meson. The
light antiquark is bound around the heavy quark by an effective
potential. The heavy quark spin decouples from the interactions as
$m_Q\to\infty$ ($m_Q$ is the mass of the heavy quark) \cite{HQ,HQ2}. 
The interactions relevant to quark spin can be treated as
perturbative correction.

Inspired by asymptotic freedom at short distance in QCD and quark
confinement at long distance, the effective potential between
quark and antiquark in meson can be taken as a combination of a
Coulomb term and a linear confining term. Such a potential will be
consistent with perturbative QCD at short distance, it can also
generate quark confinement at long distance \cite{GI,Cornell}.

The parameters in the effective potential can be constrained by
comparing the eigenvalues of the bound state wave equation with
the masses of the relevant bound states measured in experiment.
Recently the decay constants of $D$ and $D_s$ mesons have been
measured by CLEO \cite{CLEO1,CLEO2,CLEO3}, Belle \cite{Belle} and
{\it BABAR} \cite{BABAR} Collaborations. The measured values of
$D$ and $D_s$ mesons' decay constants $f_D$ and $f_{D_s}$ can give
further information about the interactions within the heavy-light
quark-antiquark system. In this paper, with the recently measured
decay constants $f_D$ and $f_{D_s}$ available, we revisit the study of the bound
states of $B$, $B_s$, $D$ and $D_s$ mesons in the relativistic
potential model.  We solve the relativistic version of
Schr\"{o}dinger equation for the bound state wave function of
heavy-light quark-antiquark meson system.  The decay constants $f_D$ and $f_{D_s}$ can be used as a further
constriant on the parameters in the potential model.  The obtained masses and decay
constants of $B$ and $D$ mesons can be well consistent with the
values measured in experiments. Then the wave functions obtained here can be more reliable
then ever.  It can be useful for studying $B$ and $D$ decays, where the momentum-distribution of the quarks is
needed.

Although the bound states of heavy mesons have been studied with
the relativistic potential model in the literature several years
before, these works need to be improved with the recent
experimental data of the decay constants of $D$ and $D_s$ mesons
available. The decay constants and bound state masses are
calculated in Refs.\cite{CCCN,PNP,CF}, where Richardson potential \cite{Ri} was taken,
here the potential we considered is different from theirs.
In addition, with the experimental values of the decay constants $f_D$ and $f_{D_s}$ available
recently, the parameters in the potential can be constrained more stringently.
Therefore our prediction on the decay constants
for $B$ and $D$ mesons are quite different from previous
predictions in the relativistic potential model.

The paper is organized as follows. In section II, we solve the
relativistic wave equation for the heavy-light quark-antiquark
system. Section III gives the decay constant in terms of the wave
function.  In section IV the QCD-inspired potential is presented.
Section V is devoted to the numerical result and discussion.
Section VI is a brief summary.

\section*{II The relativistic wave equation for heavy-light
system and the solvement}

The $B$ and $D$ mesons are assumed to be approximately described
in terms of heavy-light valence-quark configurations in the
rest-frame of the mesons. The effective potential is one-gluon
exchange dominant at short distances and a linear confinement at
long distances \cite{GI}. The equation describing the bound state
wave functions is a Schr\"{o}dinger-type wave equation with
relativistic dynamics \cite{GI,Ce,CCCN,PNP,CF}
\begin{eqnarray}
&&\left [\sqrt{-\hbar^2 \nabla^2_1+m_1^2}+\sqrt{-\hbar^2
\nabla^2_2+m_2^2}+V(r) \right ]\psi(\vec{r})\nonumber\\
&& \hspace{1cm}=E\psi(\vec{r}), \label{e1}
\end{eqnarray}
where $\vec{r}=\vec{x}_1-\vec{x}_2$ is the displacement of the
light antiquark from the heavy quark, and $\vec{x}_1$ and
$\vec{x}_2$ are the coordinates of the light and heavy quarks
respectively. The operators $\nabla_{1}$ and $\nabla_{2}$ are the
gradient operators relevant to the coordinates of $\vec{x}_1$ and
$\vec{x}_2$. $m_1$ is the mass of the light antiquark, and $m_2$
the mass of the heavy quark. $V(r)$ is the effective potential of
strong interaction between heavy and light quark-antiquark. In the
rest frame of the bound state system, the eigenvalues in the wave
equation will be the masses of the series bound states.

The wave function can be expressed in terms of spectrum
integration
\begin{eqnarray}
\psi(\vec{r})&=&\int d^3r^{\prime} \delta
^3(\vec{r}-\vec{r}^{\;\prime})\psi(\vec{r}^{\;\prime})\nonumber\\
&=&\int d^3r^{\prime}\int \frac{d^3k}{(2\pi \hbar)^3}
e^{i\vec{k}\cdot(\vec{r}-\vec{r}^{\;\prime})/\hbar}\psi(\vec{r}^{\;\prime}).
\label{e2}
\end{eqnarray}
Substitute Eq.(\ref{e2}) into Eq.(\ref{e1}), the wave equation
becomes
\begin{eqnarray}
&&\int\frac{d^3k}{(2\pi \hbar)^3}d^3r^{\prime} (\sqrt{k^2
+m_1^2}+\sqrt{k^2+m_2^2} \;
)\nonumber\\
&&\;\;\;\times e^{i\vec{k}\cdot(\vec{r}-\vec{r}^{\;\prime})/\hbar}
\psi(\vec{r}^{\;\prime})=(E-V(r))\psi(\vec{r}). \label{e3}
\end{eqnarray}
The exponential $e^{i\vec{k}\cdot\vec{r}/\hbar}$ can be decomposed
in spherical harmonics
\begin{equation}
e^{i\vec{k}\cdot\vec{r}/\hbar}=4\pi
\sum_{ln}i^lj_l(\frac{kr}{\hbar})Y^*_{ln}(\hat{k})Y_{ln}(\hat{r}),\label{e4}
\end{equation}
where $j_l(\frac{kr}{\hbar})$ is the spherical Bessel function,
$Y_{ln}(\hat{r})$ is the spherical harmonics, which satisfies the
normalization condition
\begin{equation}
\int d\Omega
Y_{l_1n_1}(\hat{r})Y_{l_2n_2}(\hat{r})=\delta_{l_1l_2}\delta_{n_1n_2}.
\end{equation}
Using the spherical harmonics decomposition of the exponential in
Eq.(\ref{e4}), and factorize the wave function into the product of
two parts: radial and angular wave functions
\begin{eqnarray}
\psi(\vec{r})=\Phi_l(r)Y_{ln}(\hat{r}),\label{e6}
\end{eqnarray}
then the wave equation of Eq.(\ref{e3}) can be transferred to be
\begin{eqnarray}
&&V(r)\Phi_l(r)+\frac{2}{\pi\hbar}\int dk \frac{k^2}{\hbar^2}\int
dr^{\prime}r^{\prime 2}(\sqrt{k^2 +m_1^2}\nonumber\\
&&\;\;\;\;+\sqrt{k^2+m_2^2} \;
)j_l(\frac{kr}{\hbar})j_l(\frac{kr^{\prime}}{\hbar})\Phi_l(r^{\prime})
=E\Phi_l(r). \label{e7}
\end{eqnarray}
For convenience later, let us define a new reduced radial wave function
$u_l(r)$ by
\begin{equation}
\Phi_l(r)=\frac{u_l(r)}{r}. \label{e8}
\end{equation}
With this definition, and for the case $l=0$ which we are
interested in this work, Eq.(\ref{e7}) becomes
\begin{eqnarray}
&&V(r)u_0(r)+\frac{2}{\pi\hbar}\int_0^\infty dk\int_0^\infty
dr^{\prime} (\sqrt{k^2 +m_1^2}\nonumber\\
&&\;+\sqrt{k^2+m_2^2}\; )
\sin(\frac{kr}{\hbar})\sin(\frac{kr^{\prime}}{\hbar})u_0(r^{\prime})
=E u_0(r), \;\;\;\;\;\label{e9}
\end{eqnarray}
where we have used the explicit expression of the spherical Bessel
function for $l=0$
\begin{equation}
j_0(x)=\frac{\sin x}{x}.
\end{equation}
Eq.(\ref{e9}) is for taking $c=1$, if recover the speed of light
appearing in the formulas, Eq.(\ref{e9}) should be
\begin{eqnarray}
&&V(r)u_0(r)+\frac{2}{\pi\hbar c}\int_0^\infty dk\int_0^\infty
dr^{\prime} (\sqrt{k^2 +m_1^2}\nonumber\\
&&\;+\sqrt{k^2+m_2^2}\; ) \sin(\frac{kr}{\hbar c
})\sin(\frac{kr^{\prime}}{\hbar c})u_0(r^{\prime}) =E u_0(r).
\;\;\;\;\;\label{e11}
\end{eqnarray}
In principle the integration over momentum $k$ in the above
equation can be performed because the wave function
$u_0(r^{\prime})$ does not depend on the momentum. However the
integration over $k$ will give a singular term for $r^\prime\to r$
in the above equation \cite{PNP}. In this work, we will take
a new step to continue to solve this equation, this
method can circumvent the appearance of the singular integral
equation.

For a bound state of two particles, when the separation between
them is large enough, the wave function will effectively vanish.
We assume such a large enough typical value for the separation
between the heavy quark and the light antiquark is $L$, then the
quark-antiquark in the bound state can be approximately treated as
if they are restricted in a limited space $0<r<L$. In the limited
space, the Fourier expansion of the reduced wave function $u_0(r)$
is
\begin{equation}
u_0(r)=\sum_{n=1}^\infty c_n \sin\left (\frac{n\pi}{L}r\right ),
\label{e12}
\end{equation}
where the expansion coefficients $c_n$ are
\begin{equation}
c_n=\frac{2}{L}\int_0^L\sin\left (\frac{n\pi}{L}r\right )u_0(r)
dr.
\end{equation}
In the limited space, the momentum $k$ should be discretized, the
integration over $k$ should be replaced by a summation, the
following substitution should be made in the wave equation
(\ref{e11})
\begin{equation}
\frac{k}{\hbar c}\to \frac{n\pi}{L}, \hspace{1.cm}
\int\frac{dk}{\hbar c}\to \frac{\pi}{L}. \label{e14}
\end{equation}
With the above replacement, and the integration over the distance
$r^\prime$ being limited within $0<r^\prime<L$, Eq.(\ref{e11})
becomes
\begin{eqnarray}
&&V(r)u_0(r)+\sum_{n}\frac{2}{L}\int_0^L
dr^{\prime} \left(\right.\sqrt{\left(\frac{n\pi\hbar c}{L}\right)^2 +m_1^2}\nonumber\\
&&\;\left.+\sqrt{\left(\frac{n\pi\hbar c}{L}\right)^2+m_2^2}\;
\right)
\sin(\frac{n\pi}{L}r)\sin(\frac{n\pi}{L}r^{\prime})u_0(r^{\prime})\nonumber\\
&&\; =E u_0(r). \;\;\;\;\;\label{e15}
\end{eqnarray}
The above equation can go back to Eq.(\ref{e11}) as $L\to \infty$.
Numerically if the value of $L$ is taken to be large enough, the
solution of this equation only slightly depends on the value of
$L$. For the parameters we take in section V, we find that the
solution of the wave equation will be stationary when $L>5\;
\mbox{fm}$.

Truncate the series of the Fourier expansion of the wave function
$u_0(r)$ as
\begin{equation}
u_0(r)=\sum_{n=1}^N c_n \sin\left (\frac{n\pi}{L}r\right ),
\label{e16}
\end{equation}
where $N$ is a large integer. Substitute this truncated expansion
into the wave equation (\ref{e15}) and simplify it, one can
finally get the equation about $c_n$
\begin{eqnarray}
&&\left(\sqrt{\left(\frac{n\pi\hbar c}{L}\right)^2
+m_1^2}+\sqrt{\left(\frac{n\pi\hbar c}{L}\right)^2+m_2^2}\;
\right)c_n\nonumber\\
&&\;+\sum_{m=1}^N\frac{2}{L}\int_0^Ldr V(r)\sin
\left(\frac{n\pi}{L}r\right)\sin
\left(\frac{m\pi}{L}r\right)c_m\nonumber\\
&&\;=Ec_n . \label{e17}
\end{eqnarray}
The above equation is just the eigenstate equation in the matrix
form. It is not difficult to solve it numerically. The eigenvalues
are the masses of the series of bound states of the heavy-light
quark-antiquark system. Once the eigen equation is solved, the
eigenvectors composed of $c_n$ can be substituted into
Eq.(\ref{e16}) to get the reduced wave function $u_0(r)$.

To get the wave function in momentum space, one can use the
Fourier transform of the wave function $\psi(\vec{r})$
\begin{equation}
\Psi(\vec{k})=\frac{1}{(2\pi\hbar c)^{3/2}}\int d^3r
e^{-i\vec{k}\cdot \vec{r}/{\hbar c}}\psi(\vec{r}). \label{e18}
\end{equation}
Separate the variable-dependence of the momentum-space wave
function as
\begin{equation}
\Psi(\vec{k})=\Psi_l(k)Y_{lm}(\theta,\phi).\label{e19}
\end{equation}
As in Eq.(\ref{e8}), we define the reduced wave function in
momentum space
\begin{equation}
\Psi_l(k)=\frac{\varphi_l(k)}{k}. \label{e20}
\end{equation}
Then using Eqs.(\ref{e4}), (\ref{e6}), (\ref{e8}), (\ref{e19}) and
(\ref{e20}), one can derive from Eq.(\ref{e18})
\begin{equation}
\varphi_l(k)=(-i)^l\sqrt{\frac{2}{\pi\hbar c}}\int_0^\infty dr
\frac{kr}{\hbar c} \;j_l(\frac{kr}{\hbar c})u_l(r).
\end{equation}
For the case $l=0$, we get
\begin{equation}
\varphi_0(k)=\sqrt{\frac{2}{\pi\hbar c}}\int_0^\infty dr
 \sin(\frac{kr}{\hbar c})u_0(r),
\end{equation}
which gives the momentum distribution of the quark and antiquark
in the rest frame of the heavy meson.

\section*{III The pseudoscalar bound state of heavy-light system and the
decay constant}

The pseudoscalar meson composed of a heavy quark and a light
antiquark $Q\bar{q}$ ($Q$ can be $b$ or $c$ quark, $q$ stands for
$u$, $d$ or $s$ quark) can be written in the meson rest frame as
follows
\begin{eqnarray}
&&|P(\vec{p}=0)\rangle =\frac{1}{\sqrt{3}}\sum_i\int
d^3k\Psi_0(k)\frac{1}{\sqrt{2}}[
b_Q^{i+}(\vec{k},\uparrow)d_q^{i+}(-\vec{k},\downarrow)
\nonumber\\
&&\hspace{1.cm}-b_Q^{i+}(\vec{k},\downarrow)d_q^{i+}(-\vec{k},\uparrow)]|0\rangle
,\label{e23}
\end{eqnarray}
where $i$ is the color index. The factor
$1/\sqrt{3}$ is the normalization factor for color indices, and
$1/\sqrt{2}$ the normalization factor for spin indices.

The normalization of the meson state is
\begin{equation}
\langle P(\vec{p}_1)|P(\vec{p}_2)\rangle =(2\pi)^3 2E
\delta^3(\vec{p}_1-\vec{p}_2), \label{e24}
\end{equation}
where $E$ is the energy of the meson.

Substituting Eq.(\ref{e23}) into Eq.(\ref{e24}), we can finally
get the normalization condition of the wave function in momentum
space
\begin{equation}
\int d^3k |\Psi_0(k)|^2=(2\pi)^3 2E. \label{norm}
\end{equation}

The decay constant of a pseudoscalar is defined by the
hadron-to-vacuum matrix element of the axial current
\begin{equation}
\langle 0|\bar{q}\gamma_\mu\gamma_5 Q|P(p)\rangle =if_P\; p_\mu .
\end{equation}
Substituting the meson state of Eq.(\ref{e23}) into the above
equation in the rest frame, and contracting the quark (antiquark)
creation operators in the meson state with quark (antiquark )
annihilation operators in the quark field of the axial current, we
can get the expression of the pseudoscalar decay constant
\begin{eqnarray}
f_P&=&\sqrt{\frac{3}{2}}\frac{1}{2\pi^2}\frac{1}{m_P}\int_0^\infty
dk |\vec{k}|^2 \Psi_0(k)\nonumber\\&&
\times\frac{(E_q+m_q)(E_Q+m_Q)-|\vec{k}|^2}{\sqrt{E_qE_Q(E_q+m_q)(E_Q+m_Q)}},
\end{eqnarray}
where $E_Q$ and $E_q$ are the energy of the heavy and light
quarks. To be consistent with the wave equation, here both heavy
and light quarks are taken to be on-shell. We assume that the
decays of the heavy meson can be approximately described in terms
of on-shell valence quarks, although the sum of the four-momenta
of the valence quarks are not equal to that of the meson because
of the existence of the color field within the hadron which can
carry both energy and momentum.

We would like to mention that the leptonic decay of pseudoscalar
mesons was considered several decades ago in a different method by assuming the
coupling of meson with quark-antiquark pair \cite{yu}.

\section*{IV The QCD-inspired potential}

The potential of strong interaction between the heavy quark and
light antiquark is taken as a combination of a Coulomb term and a
linear confining term inspired by QCD \cite{GI,Cornell}
\begin{equation}
V(r)=-\frac{4}{3}\frac{\alpha_s(r)}{r}+b\;r+c.
\end{equation}
The first term is the Coulomb term, which is consistent with
one-gluon-exchange contribution for short distance calculated in
perturbative QCD. The second term is the linear-confinement term,
which generates confinement in long distance. The third term is a
phenomenological constant, which is needed to reproduce the
correct masses for heavy-light meson system.

The running coupling constant $\alpha_s(Q^2)$ in momentum space
with $N_f$ quark flavors at large values of $Q^2$, calculated in
lowest-order QCD, is
\begin{equation}
\alpha_s(Q^2)=\frac{12\pi}{(33-2N_f)\ln
(Q^2/\Lambda^2)}.\label{e29}
\end{equation}
This behavior of the strong coupling can be parameterized in a
simpler form which can be conveniently transformed into the
$r$-space \cite{GI}
\begin{equation}
\alpha_s(Q^2)=\sum_i\alpha_i e^{-Q^2/4\gamma_i^2}, \label{e30}
\end{equation}
where $\alpha_i$ are free parameters chosen to fit the behavior of
$\alpha_s(Q^2)$ given by perturbative QCD (Eq.(\ref{e29})). As
$Q^2\to\Lambda^2$, the coupling $\alpha_s(Q^2)$ diverges, which is
believed to be a signal of confinement. However, as
$Q^2\to\Lambda^2$, perturbative QCD can not apply, the behavior of
$\alpha_s(Q^2)$ at small $Q^2$ given in Eq.(\ref{e29}) cannot be
the exact prediction of QCD. One can make other choice for the
behavior of the strong coupling at small momentum transfer. As in
Ref. \cite{GI}, we assume that the coupling $\alpha_s$ saturate as
a critical value $\alpha_s^{\rm critical}$, where $\alpha_s^{\rm
critical}=\sum_i \alpha_i$. In practice, only several $\alpha_i$
are needed to be non-zero, which can fit the behavior of
$\alpha_s(Q^2)$ well at perturbative region, deviation only occurs
at small $Q^2$.

\begin{figure}[h]
\begin{center}
\scalebox{0.5}{\epsfig{file=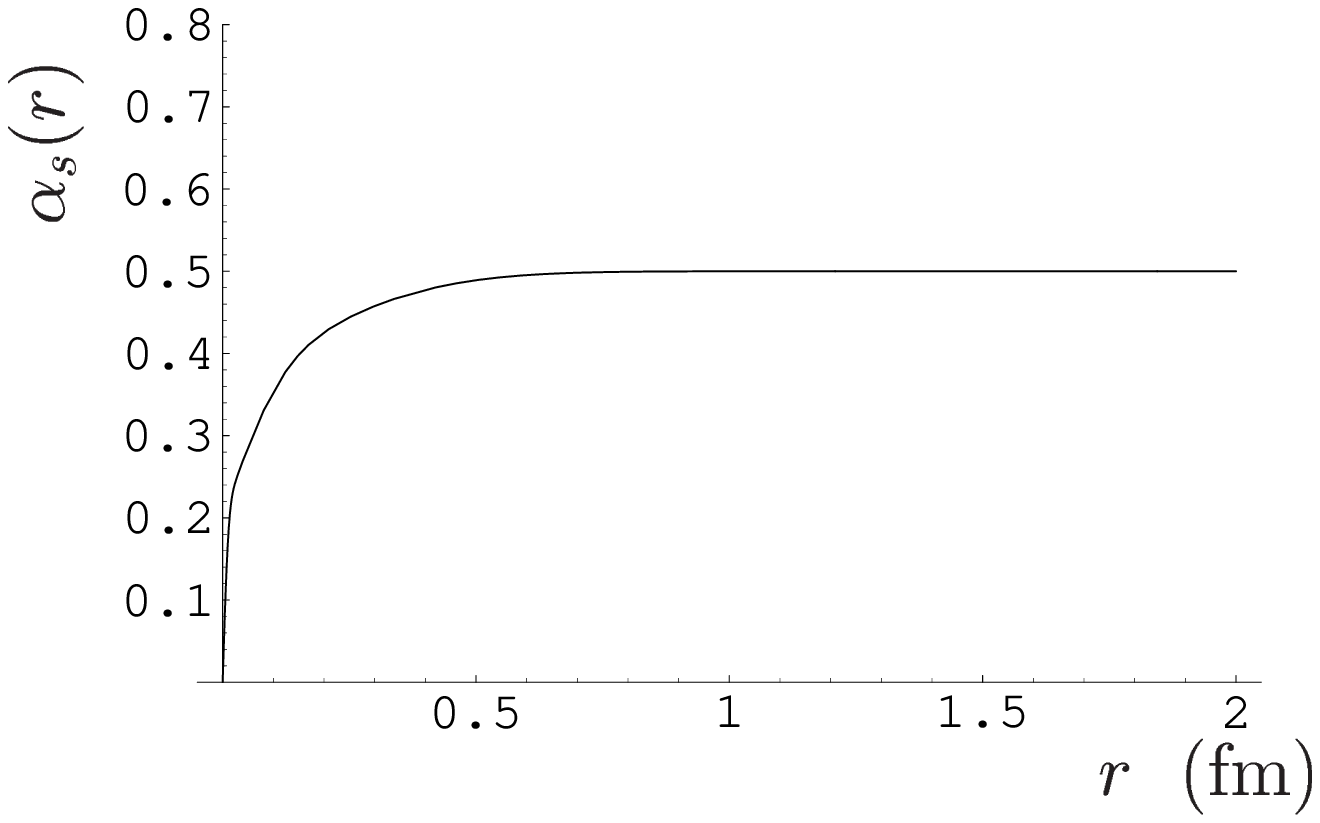}} \caption{The behavior of
$\alpha_s(r)$, with the parameters $\alpha_1=0.15$,
$\alpha_2=0.15$, $\alpha_3=0.20$, and $\gamma_1=1/2$,
$\gamma_2=\sqrt{10}/2$, $\gamma_2=\sqrt{1000}/2$.} \label{fig1}
\end{center}
\end{figure}

The transformation of $\alpha_s(Q^2)$ by using Eq. (\ref{e30})
instead of Eq. (\ref{e29}) is \cite{GI}
\begin{equation}
\alpha_s(r)=\sum_i\alpha_i\frac{2}{\sqrt{\pi}}\int_0^{\gamma_i
r}e^{-x^2}dx .
\end{equation}
Fig.\ref{fig1} is the behavior of $\alpha_s(r)$ with the
parameters $\alpha_1=0.15$, $\alpha_2=0.15$, $\alpha_3=0.20$, and
$\gamma_1=1/2$, $\gamma_2=\sqrt{10}/2$, $\gamma_2=\sqrt{1000}/2$,
which is relevant to the critical value $\alpha_s^{\rm
critical}=0.5$.

\section*{V Numerical result and discussion}

The parameters are selected by comparing the predicted heavy meson mass with the experimental
data. The recently measured values of the decay constants of $D$ and $D_s$
mesons can give a further constraint on the parameters. The parameters which we finally obtain are
\begin{eqnarray}
&& b=0.10\;{\rm GeV}^2,\;\;\; c=-0.19\; {\rm GeV}^2,\nonumber\\
&& m_b=4.98\; {\rm GeV},\;\;\; m_c=1.54\; {\rm GeV},\nonumber\\
&&m_s=0.30\;{\rm GeV},\;\;\;  m_u=m_d=0.08\; {\rm GeV},\nonumber\\
&& \alpha_s^{\rm critical}=0.5,
\end{eqnarray}
and $L=10$ fm, $N=100$.

The masses and decay constants of the $B$, $B_s$, $D$ and $D_s$ mesons calculated with the above parameters will be
given in the following. The
masses are given in Table \ref{t1}. Here we do not consider the contribution of spin-dependent
interactions in our calculation, it may give errors about $100\sim 200{\rm MeV}$ for
the masses. Varying the parameters may also give errors to the
numerical values. We estimate the combination of both the errors
can be about $7\%$ for $B$ and $B_s$ mesons, and $10\%$ for $D$
and $D_s$ mesons.

%\begin{table}\caption{Masses of pseudoscalar heavy mesons calculated by solving
 %the wave equation, and the comparison with experimental data. The data is quoted
 %from the Particle Data Group \cite{PDG}.} \label{t1}
%\begin{tabular}{|c|c|c|}\hline
%          & this work & Exp. \\ \hline
%$m_B$     & 5.25 GeV  & $5279.17\pm 0.29\; {\rm MeV}$ \\ \hline
%$m_{B_s}$ & 5.34 GeV  & $5366.3\pm 0.6 \;{\rm MeV}$   \\ \hline
%$m_D$     & 1.86 GeV  & $1869.6\pm 0.16\; {\rm MeV}$  \\ \hline
%$m_{D_s}$ & 1.96 GeV  & $1968.47\pm 0.33 \;{\rm MeV}$  \\ \hline
%\end{tabular}
%\end{table}
%\begin{widetext}
\begin{table}[h]
\begin{center}
\caption{Masses of pseudoscalar heavy mesons calculated by solving
 the wave equation, and the comparison with experimental data. The data is quoted
 from the Particle Data Group \cite{PDG}.}
 \label{t1}
\begin{tabular}{|c|c|c|c|c|}\hline
          & $m_B$ & $m_{B_s}$ & $m_D$ & $m_{D_s}$ \\ \hline
$\begin{array}{c}{\rm this\; work}\\{\rm(GeV)}\end{array}$ &
$5.25\pm 0.37$
& $5.34\pm 0.37$ & $1.86\pm 0.19$ & $1.96\pm 0.20$  \\
\hline $\begin{array}{c}{\rm Exp.}\\{\rm(MeV)}\end{array}$ &
$\begin{array}{c} 5279.17\\ \pm 0.29 \end{array}$& $\begin{array}{c} 5366.3\\ \pm 0.6\end{array}$ & $\begin{array}{c} 1869.6\\ \pm 0.16 \end{array}$ & $\begin{array}{c} 1968.47\\ \pm 0.33 \end{array}$\\
\hline
\end{tabular}
\end{center}
\end{table}
%\end{widetext}

The decay constants obtained are
\begin{eqnarray}
&&f_B=198\pm 14\;{\rm MeV},\hspace{0.3cm} f_{B_s}=237\pm 17\;{\rm MeV},\nonumber\\
&&f_D=208\pm 21\;{\rm MeV},\hspace{0.3cm} f_{D_s}=256\pm 26\;{\rm
MeV}.
\end{eqnarray}
The comparison of the decay constants obtained in this work with
experimental data are given in Table \ref{t2}. Both the masses and
decay constants obtained in this work can be well
consistent with experiment. For $f_B$ and $f_{B_s}$, there are
still no precise measured values in experiments yet. Our prediction can be tested
in experiment in the future.

The leptonic decay rates of $B$ meson relevant to
the decay constant $f_B$ obtained in this work are
\begin{eqnarray}
&&Br(B^+\to e^+\nu_e)=(1.11\pm 0.26)\times 10^{-11},\\
&&Br(B^+\to \mu^+\nu_\mu)=(4.7\pm 1.1)\times 10^{-7},\\
&&Br(B^+\to \tau^+\nu_\tau)=(1.1\pm 0.2)\times 10^{-4},
\end{eqnarray}
where the errors are mainly caused by the uncertainties of the decay constant $f_B$ and the CKM matrix element $V_{ub}$. The value of $|V_{ub}|$ is quoted from PDG \cite{PDG}
$$ |V_{ub}|=(3.93\pm 0.36)\times 10^{-3}.$$
At present the branching ratio of $Br(B^+\to \tau^+\nu_\tau)$ has been measured in experiment. The results still suffer from large uncertainties.  The measured value of Belle collaboration is $Br(B^+\to \tau^+\nu_\tau)=(1.79^ {+0.56+0.46}_{-0.49-0.51})\times 10^{-4} $ \cite{Belle-tau}, while the values of {\it BABAR}  collaboration are $Br(B^+\to \tau^+\nu_\tau)=(0.9\pm 0.6\pm 0.1)\times 10^{-4} $ \cite{BaBar-tau1} and $(1.8^{+0.9}_{-0.8}\pm 0.4\pm 0.2)\times 10^{-4} $ \cite{BaBar-tau2}.  The combined result of {\it BABAR} collaboration is $Br(B^+\to \tau^+\nu_\tau)=(1.2\pm 0.4\pm 0.3\pm 0.2)\times 10^{-4}$ \cite{Schwartz}.
Considering the large uncertainties of the experimental results, our predictied branching ratio of the decay mode $B^+\to \tau^+\nu_{\tau}$ is cosistent with the experimental data.

A super B factory will come into operation with the designed peak luminosity in excess of
$10^{36}\;\rm{cm}^{-2}\rm{S}^{-1}$ at the $\Upsilon (4s)$ resonance in the next half decades
\cite{superB}. The integrated luminosity of $75\;\rm{ab}^{-1}$ would be collected in five years
of data taking. Then the branching ratios of $B^+\to \tau^+\nu_\tau$ and $\mu^+\nu_\mu$ can be measured
at the Super{\it B} factory with precisions of up to 4\% and 5\% respectively \cite{superB}. Taking the
value of the CKM matrix element $|V_{ub}|$ as input, the decay constant of $f_B$ can be obtained at super{\it B}.

The decay constants are also compared with previous theoretical results in Table \ref{t3}.
Our predictions for the decay constants are quite different from previous results calculated
in the relativistic potential model. For $f_D$ and $f_{D_s}$, our results are larger than that
in Ref \cite{PNP}, while our results for $f_B$ and $f_{B_s}$ are smaller than theirs. \footnote{The definition of the decay constant in Ref.\cite{PNP} is different from that in the current work by a factor
$\sqrt{2}$. The quoted results of Ref. \cite{PNP} in Table \ref{t3} have been compensated by this factor.}

%\begin{widetext}
\begin{table}[h]
\begin{center}
\caption{Decay constants of pseudoscalar heavy mesons calculated
in this work, and the comparison with experimental data. All
values are in units of MeV.}
 \label{t2}
\begin{tabular}{|c|c|c|c|c|}\hline
          & $f_B$ & $f_{B_s}$ & $f_D$ & $f_{D_s}$ \\ \hline
this work & $198\pm 14$ & $237\pm 17$ & $208\pm 21$ & $256\pm 26$\\ \hline
Exp. \cite{CLEO1,HFAG}&$-$&$-$& $205.8\pm 8.5\pm 2.5$
    & $254.6\pm 5.9$ \\
\hline
\end{tabular}
\end{center}
\end{table}
%\end{widetext}

\begin{table}[h]
\begin{center}
\caption{The comparison of the decay constants calculated in this
work with other theoretical results. All values are in units of
MeV.}
 \label{t3}
\begin{tabular}{|c|c|c|c|c|}\hline\hline
          & $f_B$ & $f_{B_s}$ & $f_D$ & $f_{D_s}$ \\ \hline
this work & $198\pm 14$ & $237\pm 17$ & $208\pm 21$ & $256\pm 26$
\\  \hline
Ref.\cite{PNP}$^a$ & $230\pm 35$& $245\pm 37$&$182\pm 27$ &
$199\pm 30$
\\  \hline
Ref.\cite{Nar}$^b$ &$203\pm 23$ & $236\pm 30$& $205\pm 20$&
$235\pm 24$
\\  \hline
Ref.\cite{PS}$^b$ &$206\pm 20$ & $-$& $195\pm 20$& $-$
\\  \hline
Ref.\cite{WKY}$^c$ & $-$ & $-$ & $\begin{array}{l} 200\\230
\end{array}$ & $\begin{array}{l} 221\\270
\end{array}$ \\  \hline
Ref.\cite{duality}$^d$ &  $-$ & $-$ & $177\pm 21$& $205\pm 22$
\\  \hline
Ref.\cite{WYW}$^e$ & $193$&$195$ & $238$ & $241$
\\ \hline
Ref.\cite{CKWN}$^e$ & $196\pm 29$&$216\pm 32$ & $230\pm 25$ &
$248\pm 27$
\\ \hline
Ref.\cite{EFG}$^f$ & $189$&$218$ & $234$ & $268$
\\ \hline
Ref.\cite{GW}$^g$ & $\begin{array}{r}210\pm 11.4\\ \pm
5.7\end{array}$&$-$ & $-$ & $-$
\\ \hline
Ref.\cite{Au}$^h$  &$-$ & $-$& $\begin{array}{r}201\pm 3\\ \pm
17\end{array}$& $\begin{array}{r}249\pm 3\\ \pm 16\end{array}$
\\  \hline
Ref.\cite{Ber}$^h$  &$195\pm 11$ & $243\pm 11$& $207\pm 11$&
$249\pm 11$
\\  \hline
Ref.\cite{Fo}$^h$  &$-$&  $-$  &$207\pm 4$ & $241\pm 3$ \\
\hline
Ref.\cite{HPQCD}$^h$  &$-$&  $-$  &$-$ & $248.0\pm 2.5$ \\
\hline\hline
\end{tabular}\\[2pt]
\flushleft{ $a$ Relativistic potential model.\\
 $b$ QCD sum rule.\\
 $c$ Light front quark model.\\
 $d$ Finite energy sum rules.\\
 $e$ Quark model based on Bethe-Salpeter equation.\\
 $f$ Relativistic constituent Quark Model.\\
 $g$ Derived from result of Lattice QCD.\\
$h$ Lattice QCD}
\end{center}
\end{table}
% \vspace{0.5cm}

\begin{figure}[ht]
\begin{center}
\scalebox{0.6}{\epsfig{file=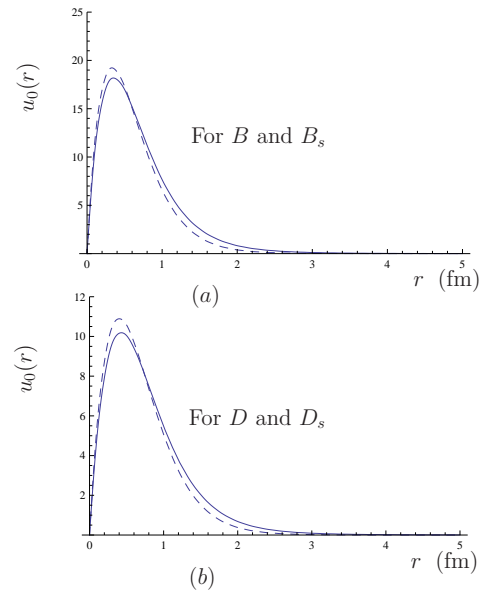}} \caption{The reduced wave
function $u_0(r)$ in coordinate space. (a) is for $B$ and $B_s$
mesons. The solid curve is for the wave function of $B$ meson, the
dashed one is for $B_s$. (b) is for $D$ and $D_s$ mesons. The
solid curve is for $D$, and the dashed one for $D_s$.}
\label{fig2}
\end{center}
\end{figure}

\begin{figure}[ht]
\begin{center}
\scalebox{0.6}{\epsfig{file=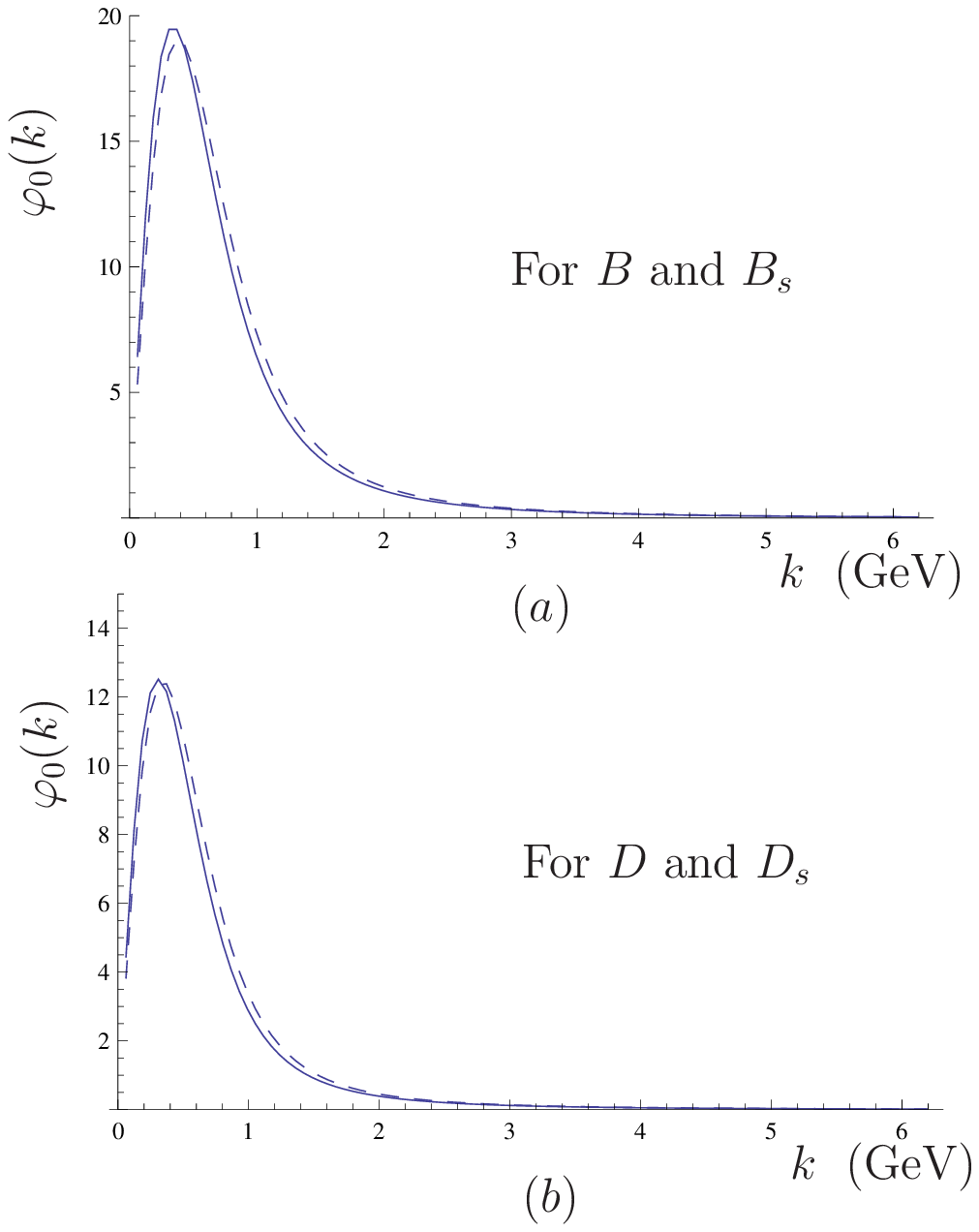}} \caption{The reduced wave
function $\varphi_0(k)$ in momentum space. (a) is for $B$ and
$B_s$ mesons. The solid curve is for $B$ meson, the dashed one is
for $B_s$. (b) is for $D$ and $D_s$ mesons. The solid curve is for
$D$, and the dashed one for $D_s$.} \label{fig3}
\end{center}
\end{figure}

The reduced wave functions in coordinate and momentum spaces (Eqs.
(\ref{e8}) and (\ref{e20}) are depicted in Figs. \ref{fig2} and
\ref{fig3}. The wave function squared $|u_l(r)|^2$ is the
possibility density distributed along the quark-antiquark distance
$r$. The curves in Fig. \ref{fig2} show that the most
probable distribution of the quarks occurs at the distance 0.4 fm
between the quark and antiquark in both $B$ and $D$
mesons. The possibility density vanishes as the distance larger
than 2 fm. The mean square root of the distance is about $0.5\sim
0.7$ fm.

The numerical solution of the wave function in momentum space is given in Fig.\ref{fig3},
which shows that the peak of the momentum-distribution of the quarks in the heavy meson is at
about 0.4 GeV. The reduced wave function can be fitted with the analytical form as suggested in Ref.\cite{CF}
\begin{equation}
\varphi_0(k)=4\pi\sqrt{m_H \alpha^3}k e^{-\alpha k},
\end{equation}
where $m_H$ is the heavy meson mass. The factor $4\pi\sqrt{m_H \alpha^3}$ is the normalization
factor due to the normalization condition in Eq.(\ref{norm}). Note that the wave function for $B$ and/or $D$ meson is $\Psi_0(k)=\varphi_0(k)/k$.  Our numerical solution gives $\alpha=3.0\;{\rm GeV}^{-1}$, $2.6 \;{\rm GeV}^{-1}$,
$3.4\;{\rm GeV}^{-1}$, and $3.2\;{\rm GeV}^{-1}$
for $B$, $B_s$, $D$ and $D_s$ mesons, respectively.

With the constraint of the measured values of $f_D$ and $f_{D_s}$ considered, the wave functions obtained here can be
more reliable than before, which should be useful in studying the decays of the $B$ and $D$ mesons. The application of
the wave functions in studying the heavy meson decays deserves a separate work.

\section*{VI Summary}
The wave functions and decay constants of $B$ and $D$ mesons are
revisited in the relativistic potential model. The parameters in the potential model are
further constrained with the experimental values of $f_D$ and $f_{D_s}$ available. The
masses and decay constants of the heavy mesons are obtained, which can be well consistent with
the current experimental data. The wave functions both in coordinate and
momentum spaces are obtained. The wave functions obtained here can be useful for
studying heavy meson decays.

%%%%%%%%%%%%%%%%%%%%%%%%%%%%%%%%%%%%%%%%%%%%%%%%%%%%%%%%%%%%%%%%%%%%%%%%%
% ACKNOWLEDGMENTS
%%%%%%%%%%%%%%%%%%%%%%%%%%%%%%%%%%%%%%%%%%%%%%%%%%%%%%%%%%%%%%%%%%%%%%%%%

\section*{Acknowledgments} This work is supported in part by the
National Natural Science Foundation of China under contracts Nos.
10575108, 10975077, 10735080, and by the Fundamental Research Funds for the
Central Universities No. 65030021.

%%%%%%%%%%%%%%%%%%%%%%%%%%%%%%%%%%%%%%%%%%%%%%%%%%%%%%%%%%%%%%%%%%%%%%%%%
% BIBLIOGRAPHY
%%%%%%%%%%%%%%%%%%%%%%%%%%%%%%%%%%%%%%%%%%%%%%%%%%%%%%%%%%%%%%%%%%%%%%%%
%\begin{thebibliography}{99}

%\bibliographystyle{h-physrev2-original}   %

\end{document}